\documentclass[10pt]{article}
\usepackage{epsfig}
\usepackage{url}
\usepackage{algorithm}
\usepackage{algorithmic}

\begin{document}

\title{Increasing precision of uniform pseudorandom number generators}
\author{Vadim Demchik\thanks{Email: vadimdi@yahoo.com}~ and Alexey Gulov\thanks{Email: alexey.gulov@gmail.com}\\~\\~
 ~{\small \sl Dnipropetrovsk National University, Dnipropetrovsk, Ukraine}
}

\maketitle

\begin{abstract}
A general method to produce uniformly distributed pseudorandom numbers with
extended precision by combining two pseudorandom numbers with lower precision
is proposed. In particular, this method can be used for pseudorandom number
generation with extended precision on graphics processing units (GPU),
where the performance of single and double precision operations can vary
significantly.
\end{abstract}


\section{Introduction}\label{algo:f}
Rapid development of computers and computing methods causes new requirements
to computational algorithms. Significant difference in the performance of
computational systems using single and double precision causes search for
new methods of optimization of existing algorithms. One of the most popular
classes of algorithms is pseudorandom number generators (PRNGs) with the performance
and the statistical properties, which conversely affect on many numerical methods.

General purpose computing on graphics processing units is an example of an alternative
computing platform with special architecture, which has become widespread in recent years.
A characteristic feature of such hardware is significantly different performance in
applications with single and double precision. The architecture of modern GPUs is
designed to obtain the best performance in floating point operations with single precision.
Top-end GPUs have performance on the double precision arithmetics as a half of the performance
with single precision \cite{NVIDIA:2014}. At the same time middle and low-end GPUs show much
poorer performance in double precision floating point operations. In particular, the
ratio of single-to-double precision performance even reaches 24 (for NVIDIA Tesla K10 GPU).

A problem of pseudorandom numbers (PRNs) generation with extended precision from several PRNs
with reduced precision is not a new problem. In particular, it was discussed in the papers
\cite{Doornik:2007}, \cite{TestU01:2007}. However, insufficient attention was paid to the
question about the distribution of resulting PRNs.

In this paper we propose a general method for generation of uniformly distributed
PRNs with extended precision, which is based on a regular pseudorandom number generation
algorithm. The key feature of the proposed method is to strictly preserve the uniformity
of the distribution of PRNs with extended precision.

\section{Description of the method}
According to the IEEE 754 standard the fractional part of floating point double precision number
is stored in the lowest 52 bits (in the lowest 23 bits for single precision number). So, it is
possible to generate PRNs in interval $[0;1)$ by combining two uniformly distributed 32-bit
unsigned integer PRNs or with two PRNs with single precision.

The simplest way to construct PRN $z$ with extended precision is the following:
\begin{eqnarray}\label{extendedPRNG}
z=x_1+kx_2,
\end{eqnarray}
where $x_1$ and $x_2$ are initial PRNs with $w$-bit precision, $k=2^{-w}$. Obviously, $z$ is
$2w$-bit precision floating point number. If computational unit supports the IEEE 754 double
precision arithmetics and $w\ge 26$, then $z$ can be reduced to a full double precision number.

Let us discuss the general case where $x_1\in[a_0;a]$ and $x_2\in[b_0;b]$ are initial PRNs.
The difference in the range of values may be caused by different PRNGs producing $x_1$ and $x_2$ numbers.
The probability density functions (PDF) $f_1$ and $f_2$ for $x_1$ and $x_2$, correspondingly, are constant:
\begin{eqnarray}
f_1(a_0<x_1<a)=1/(a-a_0),~~~f_1=0~\rm{otherwise},\\\nonumber
f_2(b_0<x_2<b)=1/(b-b_0),~~~f_2=0~\rm{otherwise}.
\end{eqnarray}

To determine the interval where $z$ is uniformly distributed (the PDF $f(z)$ is constant) we study
intersections of lines $z=\mathrm{const}$ with the rectangular area $x_1\in[a_0;a]$, $x_2\in[b_0;b]$.
Since the area corresponds to uniform probability, we have to select the intersections with equal
lengths. Taking into account that $k\ll 1$, we find the parallelogram bounded by the lines crossing
points $x_1=a_0$, $x_2=b$ and $x_1=a$, $x_2=b_0$ (the region $ABDE$ on Fig.1). Therefore, the PDF
$f(z)$ is constant only in the interval
\begin{eqnarray}\label{zinterval}
z\in[a_0+kb;a+kb_0].
\end{eqnarray}
In this regard, we propose accept-reject method to select $(x_1,x_2)$ pairs, which form $z$ and lie
in the desired interval.

The rejected values belong to intervals $x_1<a_0+k(b-b_0)$ and $x_1>a-k (b-b_0)$.
\begin{figure}
\begin{center}
\ifx\pdftexversion\undefined
\includegraphics[bb=77 44 777 461,width=0.8\textwidth]{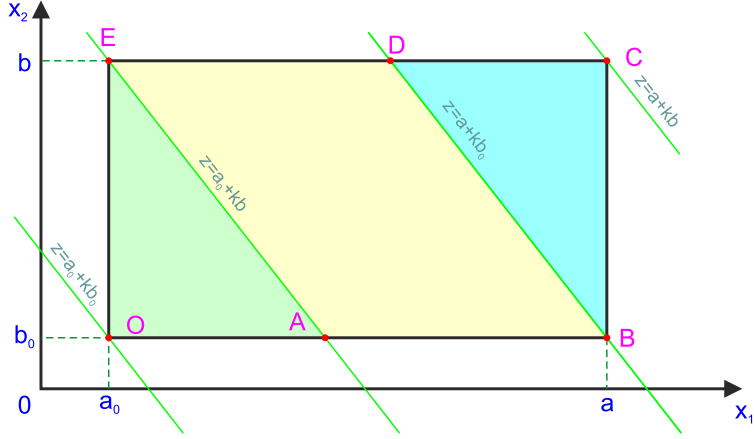}
\else
\includegraphics[width=0.8\textwidth]{uniform.eps}
\fi
\end{center}
\caption{The graphics representation of allowed region for constructed PRN $z$.}
\end{figure}
In practice, $b-b_0\simeq 1$ and $x_1$ is discrete with the step $k$, therefore the rejection area
degenerates into the boundary points of $x_1$. In this case, accept-reject condition can be formulated
as follows: {\bf if $x_1=a_0$ or $x_1=a$ then we drop such values and take next
pair $(x_1,x_2)$}.

The further step is to map the interval $z\in(a_0+kb;a+kb_0)$ to $z'\in[0;1)$. For continuous variables
\begin{eqnarray}\label{doubleconv}
z'=\frac{z-(a_0+kb)}{a-a_0-k(b-b_0)}=\frac{x_1+k x_2 - (a_0+kb)}{a-a_0-k(b-b_0)}.
\end{eqnarray}

The algorithm for generation of PRN with increased precision from two integer PRNs
with lower precision is described below.

\begin{minipage}[c]{10cm}
 \begin{algorithm}[H]
\caption{\textsc{PRNGD}($k, min, max$)}
\begin{algorithmic}
 \STATE $R1 \gets -1.0$ \hskip 0.2cm \COMMENT{initialize first PRN, double precision (DP)}
 \STATE $Rmin \gets (min+k*(max-min))$
 \STATE $Rmax \gets (max-k*(max-min))$
 \WHILE{($R1 < Rmin$ or $R1 > Rmax$)}
   \STATE $r1 \gets PRNG()$ \hskip 0.1cm \COMMENT{get first PRN, single precision (SP)}
   \STATE $R1 \gets DP(r1)$ \hskip 0.2cm \COMMENT{convert $r1$ into DP $R1$}
 \ENDWHILE
 \STATE $r2 \gets PRNG()$ \hskip 0.4cm \COMMENT{get second PRN (SP)}
 \STATE $R2 \gets DP(r2)$ \hskip 0.5cm \COMMENT{convert $r2$ into DP $R2$}
 \STATE $Z = (R1 + k * R2 - (min + k * max)) / ((max - min) * (1 - k))$
\RETURN $(Z)$
\end{algorithmic}
\end{algorithm}
\end{minipage}
\vskip 0.5cm
\noindent Here, $min$ and $max$ are the minimal and maximal values, which can be produced by
initial PRNG, $k$ is a resolution parameter of initial PRNG (see (\ref{extendedPRNG})).
It should be noted that parameters $Rmin$ and $Rmax$ as well as the most part of $Z$ are calculated
during the compilation time, so they are constants in runtime.

Considering production of the double precision PRNs from floating point PRNs with lower precision and
the accept-reject condition, we must omit the digits outside the precision of $x_1$. In this case, in
order to include $z'=0$ and exclude $z'=1$ we must map $x_1=a_0+k$, $x_2=b_0$ onto $z'=0$, and
$x_1=a-k$, $x_2=b$ onto $z'=1-k'$, where $k'=k^2$ marks the resulting precision. Thus,
Eqn.~(\ref{doubleconv}) can be substituted by
\begin{eqnarray}\label{doublefloatconvgen}
z'=\frac{
{\textrm{trunc}}\left[\frac{x_1-a_0-k}{k}\right]+x_2 - b_0
}{
{\textrm{trunc}}\left[\frac{a-a_0-2k}{k}\right]+b-b_0
}(1-k').
\end{eqnarray}
Here $\textrm{trunc}[x]$ is rounding function, returning the nearest integer value
that is not larger in magnitude than $x$. Due to the fact that the function $\textrm{trunc}[x]$ is well
optimized on the GPU as well as the fact that most of the arithmetic operations in the latter expression
are performed while compilation, Eqn.~(\ref{doublefloatconvgen}) is not resource-intensive.

In case of one production PRNG in the interval $[0,1-k]$:
\begin{eqnarray}\label{doublefloatconv}
z'=\frac{(1-k^2)\left({\textrm{trunc}}\left[\frac{x_1}{k}\right]+x_2-1\right)}
{
{\textrm{trunc}}\left[\frac{1}{k}\right]-2-k
}
,
\end{eqnarray}
where $x_1=0$ and $x_1=1-k$ are rejected.

In some cases one needs a generator, which does not produce zero values (for example,
if the PRN will be used under logarithm and manual limit of the divergence is required).
A common well-known method is to use $1-z$ instead of $z$.

Certainly, the keystone of the proposed method is the assumption that the original random numbers,
from which a new random number is constructed, are independent and uniformly distributed. In this
regard, it makes sense to use only generators which have good statistical properties (high-level
RANLUX, MRG32k3a, RANMAR, etc.)

\section{Conclusion}
We propose a general method to produce uniformly distributed PRNs with
extended precision by combining two PRNs with lower precision.
To ensure the resulting distribution is uniform the accept-reject method is used.
This scheme can be interest for the computational facilities with significantly
different performance of single and double precision arithmetics. GPU is a popular
class of such hardware.

The proposed scheme can be generalized for the case when it is necessary to combine more then two
PRNs to obtain random numbers with greater precision. However, statistical impurities of initial
PRNGs can destroy the assumed uniform distribution of resulting PRNs. This subject requires
additional study for the specific PRNG.

\end{document}